\documentclass[sigconf]{acmart}
\AtBeginDocument{%
  }

\copyrightyear{2025}
\acmYear{2025}
\setcopyright{cc}
\setcctype{by}
\acmConference[FSE Companion '25]{33rd ACM International Conference on the Foundations of Software Engineering}{June 23--28, 2025}{Trondheim, Norway}
\acmBooktitle{33rd ACM International Conference on the Foundations of Software Engineering (FSE Companion '25), June 23--28, 2025, Trondheim, Norway}
\acmDOI{10.1145/3696630.3730564}
\acmISBN{979-8-4007-1276-0/2025/06}

\usepackage{algorithmic}
\usepackage{graphicx}
\usepackage{textcomp}
\usepackage{tcolorbox}
\usepackage{todonotes}
\usepackage{url}
\usepackage{listings}
\usepackage{booktabs}
\usepackage{xspace}
\usepackage{balance}
\usepackage{makecell}

\newcommand{\approach}{\textsc{MultiMind}\xspace}
\newcommand{\repository}{{\url{https://gitlab.com/learnERC/llm-code-plugin}}\xspace}

\begin{document}

\title{\approach: A Plug-in for the Implementation of Development Tasks Aided by AI Assistants}



\author{Benedetta Donato}
\email{benedetta.donato@unimib.it}
\orcid{0009-0001-1548-4429}
\affiliation{%
  \institution{University of Milano-Bicocca}
  \city{Milano}
  \country{Italy}
}

\author{Leonardo Mariani}
\email{leonardo.mariani@unimib.it}
\orcid{0000-0001-9527-7042}
\affiliation{%
  \institution{University of Milano-Bicocca}
  \city{Milano}
  \country{Italy}
}

\author{Daniela Micucci}
\email{daniela.micucci@unimib.it}
\orcid{0000-0003-1261-2234}
\affiliation{%
  \institution{University of Milano-Bicocca}
  \city{Milano}
  \country{Italy}
}

\author{Oliviero Riganelli}
\email{oliviero.riganelli@unimib.it}
\orcid{0000-0003-2120-2894}
\affiliation{%
  \institution{University of Milano-Bicocca}
  \city{Milano}
  \country{Italy}
}

\author{Marco Somaschini}
\email{m.somaschini@campus.unimib.it}
\affiliation{%
  \institution{University of Milano-Bicocca}
  \city{Milano}
  \country{Italy}
}

\renewcommand{\shortauthors}{Donato et al.}

\begin{abstract}
The integration of AI assistants into software development workflows is rapidly evolving, shifting from automation-assisted tasks to collaborative interactions between developers and AI. Large Language Models (LLMs) have demonstrated their effectiveness in several development activities, including code completion, test case generation, and documentation production. 
However, embedding AI-assisted tasks within Integrated Development Environments (IDEs) presents significant challenges. It requires designing mechanisms to invoke AI assistants at the appropriate time, coordinate interactions with multiple assistants, process the generated outputs, and present feedback in a way that seamlessly integrates with the development workflow. To address these issues, we introduce \approach, a Visual Studio Code plug-in that streamlines the creation of AI-assisted development tasks. \approach provides a modular and extensible framework, enabling developers to cost-effectively implement and experiment with new AI-powered interactions without the need for complex IDE customizations. \approach has been tested in two use cases: one for the automatic generation of code comments and the other about the definition of AI-powered chat.
\end{abstract}




\maketitle

\section{Introduction} \label{sec:introduction}
How development tasks are accomplished is quickly changing from tasks performed by developers assisted by automation tools to tasks performed by developers in strict collaboration with AI assistants~\cite{Terragni:FutureAISE:TOSEM:2025}. In this context, an AI assistant refers to either an AI model or an AI service that supports software development activities. For instance, AI assistants can interactively suggest new statements to complete the implementations of functions written by programmers~\cite{Copilot}, they can generate test cases to validate software components~\cite{Rao:CATLM:ASE:2024}, and produce documentation to explain code~\cite{Khan:CodeCod:ASE:2023}.

AI assistants are usually empowered by deep learning models, and in particular, Large Language Models (LLMs), which have demonstrated their effectiveness with generative tasks. Indeed, LLMs promise to increase productivity, effectively completing tasks that could not be completed without exploiting the knowledge that these models embed.

The engineering of the interactions between developers and AI assistants is however still in its infancy. The knowledge about how to design and master these interactions is limited, and recent empirical studies show several important areas of improvement that must still be investigated~\cite{liang2024large}.

Researchers are actively exploring how development tasks should be redefined to exploit at best the contribution of, potentially many, AI assistants. For instance, Qian et al.~\cite{qian-etal-2024-experiential} investigated the employment of multiple LLMs assigned with different roles  in software development. Their framework, Experiential Co-Learning, enables instructor and assistant agents to iteratively refine their outputs by leveraging past task experiences, improving efficiency and reducing human intervention. Similarly, Ulfsnes et al.~\cite{ulfsnes2024transforming} examined how the integration of AI assistants, such as ChatGPT and Copilot, is transforming software development workflows. Their study highlights a shift in collaboration dynamics, where developers increasingly rely on AI tools instead of consulting colleagues, affecting team-based learning and knowledge-sharing practices.

Designing and experimenting with new tasks is however expensive. The realization of a development task assisted by AI within an IDE requires implementing mechanisms to trigger assistants at the proper time, handle the interaction with one or more assistants, process the collected results, and visualize feedback for the developers. Implementing these interactions from scratch is expensive and cumbersome for both the researchers who want to experiment with new tasks and the practitioners who want to strengthen their IDEs with new capabilities.

To address the need for rapidly developing new tasks that can be experimented and assessed within IDEs, this paper proposes \approach, which is a plug-in for Visual Studio Code~\cite{vscode} designed to facilitate the implementation of novel tasks that involve AI assistants. 

\approach consists of some key components that can be extended and customized based on the specific analysis that has to be implemented. These components include \emph{Drivers}, which manage interactions with various AI assistants, the \emph{Driver Manager}, which coordinates these interactions, \emph{Tasks}, which define the AI's objectives, the \emph{Task Manager}, which oversees the execution of tasks and ensures smooth communication between models, and \emph{Actions}, which are user-triggered operations within the VSCode interface.

We made \approach publicly available so that researchers and practitioners are encouraged to extend and experiment with our plugin. Moreover, we used the plug-in to implement development tasks that require AI assistants, such as comment generation, code quality evaluation, and code generation, demonstrating its flexibility and usefulness. Our experience shows that \approach effectively simplifies the integration of various AI assistants into development workflows, allowing for smoother and more efficient interactions. Additionally, it provides a robust platform for experimentation, enabling users to customize tasks according to their specific needs. This adaptability enhances productivity and fosters an environment for innovation in AI-assisted software development.

The paper is organized as follows. Section~\ref{sec:VisualStudioIntegration} discusses the integration of Visual Studio Code, while Section~\ref{sec:TheIDENAMEPlugin} details \approach's architecture, including its key components such as actions, tasks, and the task manager. Section~\ref{sec:Implementing AI-Assisted Analyses with Pincopallo} guides users on implementing AI-assisted analyses with \approach, and Section~\ref{sec:IDENAMEInAction} showcases practical applications of the plugin. Finally, Section~\ref{sec:RelatedWork} reviews related work, concluding with Section~\ref{sec:Conclusion}, which outlines future research directions.

\section{Visual Studio Code Integration} \label{sec:VisualStudioIntegration}
Integrated Development Environments (IDEs) are foundational tools in contemporary software engineering, enabling developers to efficiently design, implement, and maintain complex systems, supporting tasks such as code analysis, refactoring, and integration with external development tools. Microsoft Visual Studio Code (VSCode)~\cite{vscode} is one of the most sophisticated and widely adopted IDEs, because of its extensive feature set, multi-language support, and robust extensibility mechanisms. 

This section provides a technical overview of the Visual Studio Code’s architecture, its extensibility framework, and the underlying principles governing the development of plugins, forming the basis for understanding the integration of our extension. 

VSCode is an IDE for the development of applications across multiple domains, including desktop, web, cloud, and mobile environments. 
It supports a broad spectrum of programming languages, including C++, C\#, F\#, and Python, and offers comprehensive tooling such as IntelliSense~\cite{IntelliSense} for intelligent code completion, an integrated debugging environment, and built-in version control integration with Git~\cite{vscode}.

A key characteristic of VSCode is its extensibility, which is achieved through a well-defined API infrastructure that allows third-party developers to customize and augment the IDE’s capabilities. These APIs allow developers to create extensions that can add new features, integrate with other tools, and modify the existing functionality of the editor. This extensibility makes VSCode a highly versatile and powerful tool for developers, enabling them to tailor their development environment to their specific needs and workflows. The availability of these APIs has fostered the development of a rich ecosystem of plugins and extensions aimed at enhancing productivity, automating repetitive tasks, and integrating external services~\cite{vscodeAPI}.

The extensibility framework of Visual Studio is structured around the Visual Studio SDK (Software Development Kit), which provides a set of libraries, interfaces, and services~\cite{VsSdkDocs}. Developers can extend Visual Studio by leveraging two primary mechanisms: the Managed Extensibility Framework (MEF) and the Visual Studio Package (VSIX) system. MEF enables the creation of lightweight, modular enhancements by allowing developers to define and compose components dynamically, without the need for full-package deployment~\cite{MEF2020}. The VSIX format is the standard packaging mechanism for extensions, encapsulating the necessary metadata, binaries, and configuration files required for deployment.

From an architectural standpoint, Visual Studio is built upon a modular, layered framework that enables systematic interaction between extensions and core components such as the Solution Explorer, the code editor, and the debugging interface~\cite{vscode}. 

Visual Studio extensions may target different elements. For instance, editor extensions may enhance the development experience by modifying syntax highlighting and providing autocompletion; while command and menu extensions may integrate new commands, menu options, and toolbar actions.






To maintain system integrity and security, Visual Studio enforces a sandboxed execution model for extensions, restricting unauthorized modifications to core IDE functionalities. Additionally, Microsoft curates an official extension distribution platform, the Visual Studio Marketplace, which imposes a stringent validation process to ensure quality, compatibility, and security compliance~\cite{VsMarketplace}.



\section{The \approach Plugin} \label{sec:TheIDENAMEPlugin}
The \approach Plugin is an extension for Visual Studio Code (VSCode) designed to seamlessly interact with multiple Large Language Models (LLMs), to assist researchers and practitioners in creating new analyses that leverage the power of Artificial Intelligence (AI). 

\approach allows developers to configure several AI assistants, define tasks that can be completed by interacting with these models, and orchestrate the interaction among multiple models, to finally solve complex tasks.

\begin{figure}[ht]
        \centering
        \includegraphics[width=0.485\textwidth]{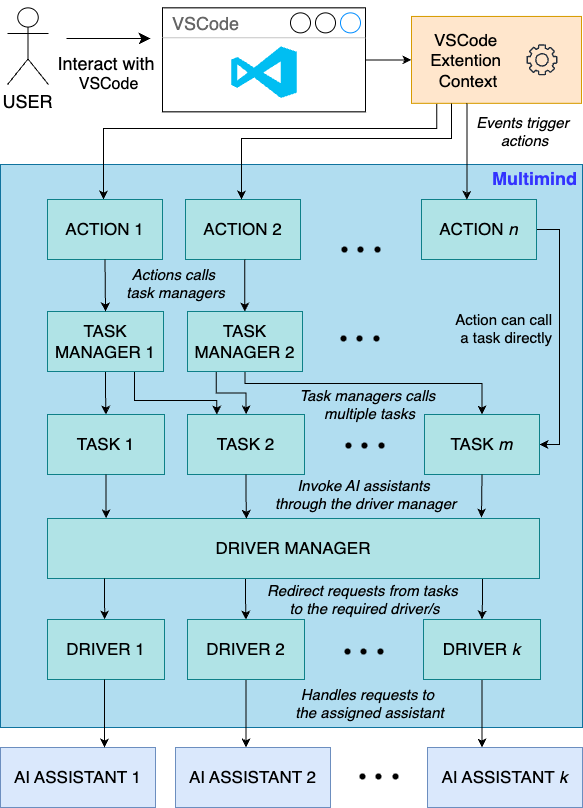} 
        \caption{Overall architecture of the \approach Plugin.} 
        \label{fig:architectural_schema}
\end{figure}

The \approach Plugin consists of the following main components (see Figure~\ref{fig:architectural_schema} for an overview) that decompose the responsibilities that characterize the execution of a task that involves interaction with one or more AIs.

\begin{itemize}
    \item \textbf{Action(s):} they are the specific events that users trigger from the VSCode interface. The response to these events typically requires the execution of a task that involves one or more AIs.  
    \item \textbf{Task Manager:} coordinates and oversees the execution of one or multiple tasks activated in response to the execution of an action, ensuring efficient information flow among tasks.
    \item \textbf{Task(s):} specific jobs executed with the contribution of AI assistants.
    \item \textbf{Driver Manager:} a central entity that manages all the Drivers and coordinates individual requests to AIs.
    \item \textbf{Driver(s):} they are components that manage the interaction with AI assistants. Each driver is configured to interact with a specific AI instance (e.g., a given version of ChatGPT or Copilot).
\end{itemize}

In the following sections, each module is described in detail, illustrating how they interact and contribute to the overall architecture of the \approach Plugin.

\subsection{Actions}

Actions represent the first entity executed when a user interacts with the plugin. For instance, if the \approach is configured to provide the capability to generate the Javadoc of a Java method when the method name is highlighted, the corresponding Action object is executed as soon as any method is highlighted. To achieve this, actions are registered to be executed as a reaction to specific GUI events generated by the IDE.  



When a single task must be executed to respond to an action, the action directly invokes the corresponding task. For instance, if the user highlights a method name and selects the \textit{Generate Javadoc} option from the context menu, the action associated with this command is triggered, subsequently invoking the task responsible for generating the documentation.  On the other hand, if the action requires the execution of multiple tasks, such as generating documentation followed by evaluating its quality, the action would call the task manager. The task manager then coordinates the execution of these tasks, ensuring they are processed in the correct order and that their results are integrated effectively before presenting the final output to the user.



Finally, actions may draw new visual objects in the IDE if necessary to effectively visualize the results. For instance, an action may open a \texttt{web component} with a chat to start a natural language conversation between the user and the AIs.


\subsection{Task Managers} 
Task Managers serve as intermediaries between Tasks and Actions, coordinating the execution of multiple tasks, and thus enabling the analysis and integration of results produced by different AI assistants.

Task Managers are particularly useful when Tasks need to be executed either in parallel or sequentially. For instance, in a scenario where one AI generates code and another verifies its correctness, the Task Manager orchestrates their interactions, ensuring a structured workflow where the second AI is activated only once the outcome of the first AI is available. Moreover, a task manager may also implement an iterative process that keeps asking for modifications to the generation AI, until the result passes the check of the verification AI. 



\approach provides a flexible framework that allows developers to create custom Task Managers tailored to specific workflows and requirements. This customization capability encourages developers to define their task execution strategies, coordinating various AI interactions based on the unique needs of their projects.

\subsection{Tasks}
Tasks are implementations of specific routines relevant to users that AI assistants must fulfill. For instance, tasks might be about recommending new code or tests, generating comments, and fixing bugs. The implementation of a task may range from a simple query to a model to complex decision-making processes. Tasks are categorized based on their level of specificity:

\begin{itemize}
    \item \textit{Defined Tasks:} These tasks follow well-structured requirements with detailed contexts. They are ideal when AI responses must adhere to strict constraints, ensuring consistency and reliability. Typically, these tasks can be completed in one step without needing further input from the user. One example is documentation generation, where the role of AI is clearly defined and contextualized.
    \item \textit{Open-ended Tasks:} These tasks offer greater flexibility, allowing users to provide broad instructions that guide AI decision-making. They are particularly useful for scenarios such as creative writing, exploratory analysis, and summarization, where creativity and adaptability are key. Often, these tasks require multiple interactions to complete effectively, as they involve a higher degree of user input and feedback. An example of this is the sidebar chat, where the AI addresses a wide range of requests, frequently without a detailed context.

\end{itemize}

Each Task can interact with the Driver Manager according to two main modes. 
\begin{itemize}
    \item \emph{Continue After First Response}: this mode privileges speed over accuracy. That is, multiple AIs are queried but the task continues its execution as soon as the first response is available, ignoring the late responses. This approach is ideal for scenarios requiring real-time feedback and interaction, ensuring minimal latency.
    \item \emph{Continue After Last Response}: this mode waits for all the queried AIs to produce a response before proceeding. 
    This ensures a comprehensive evaluation of multiple responses before making a final decision or executing further operations.
\end{itemize}

When querying the Driver Manager, Tasks can require a specific model to be used, to leverage model-specific strengths, ensuring that the best-suited AI is used for each task. 


\subsection{Driver Manager} 
The Driver Manager is the core component responsible for managing and orchestrating all Drivers within the \approach plugin. Acting as an abstraction layer, it enables seamless interaction with multiple AI assistants while easing model selection, communication, and result aggregation. 

Designed as a singleton, the Driver Manager ensures that only one instance is active, preventing redundancy and optimizing resource allocation. Its responsibilities include registering the available Drivers, configuring operational parameters, and monitoring the activity.

On a technical perspective, the Driver Manager provides two main methods for AI interaction: 
\begin{itemize} 
    \item \texttt{CallBack:} implements an asynchronous mechanism that retrieves the first available AI response while allowing other processes to continue execution. This approach reduces latency, making it useful for real-time interactions where immediate feedback is required. It implements the Continue After First Response mode.
    
    \item \texttt{FetchAll:} collects all AI responses before proceeding, ensuring that a comprehensive dataset is available for decision-making. This method is ideal for tasks requiring complete (and sometimes redundant) information before execution. It implements the Continue After Last Response mode.
\end{itemize}

These mechanisms, jointly with the possibility to select the individual AI assistants to be used to respond to a given request, allow the system to dynamically adapt its AI interaction strategy based on task-specific requirements. 


\subsection{Drivers} 
Drivers serve as the fundamental bridge between the \approach plugin and various AI assistants or external APIs. Their primary role is to facilitate seamless communication between the extension and external AI assistants, ensuring interoperability across different models. 

Each Driver instance is configured to interact with an AI assistant or model according to a user-specific configuration. The typical responsibilities of a driver include 
handling authentication and session management, formatting requests according to the expected input structure, sending requests via the appropriate communication protocol, collecting responses and transforming them into a format suitable for the plugin. 

The hierarchy of drivers is easily extensible with the possibility to configure and integrate new  AI assistants and models. The current implementation already includes several driver implementations, such as those for connection to GPT models and Gemini, providing a versatile framework that can adapt to the evolving landscape of AI technologies.




\subsection{Typical Execution in \approach}
\begin{figure*}[ht]
        \centering
        \includegraphics[width=0.98\textwidth]{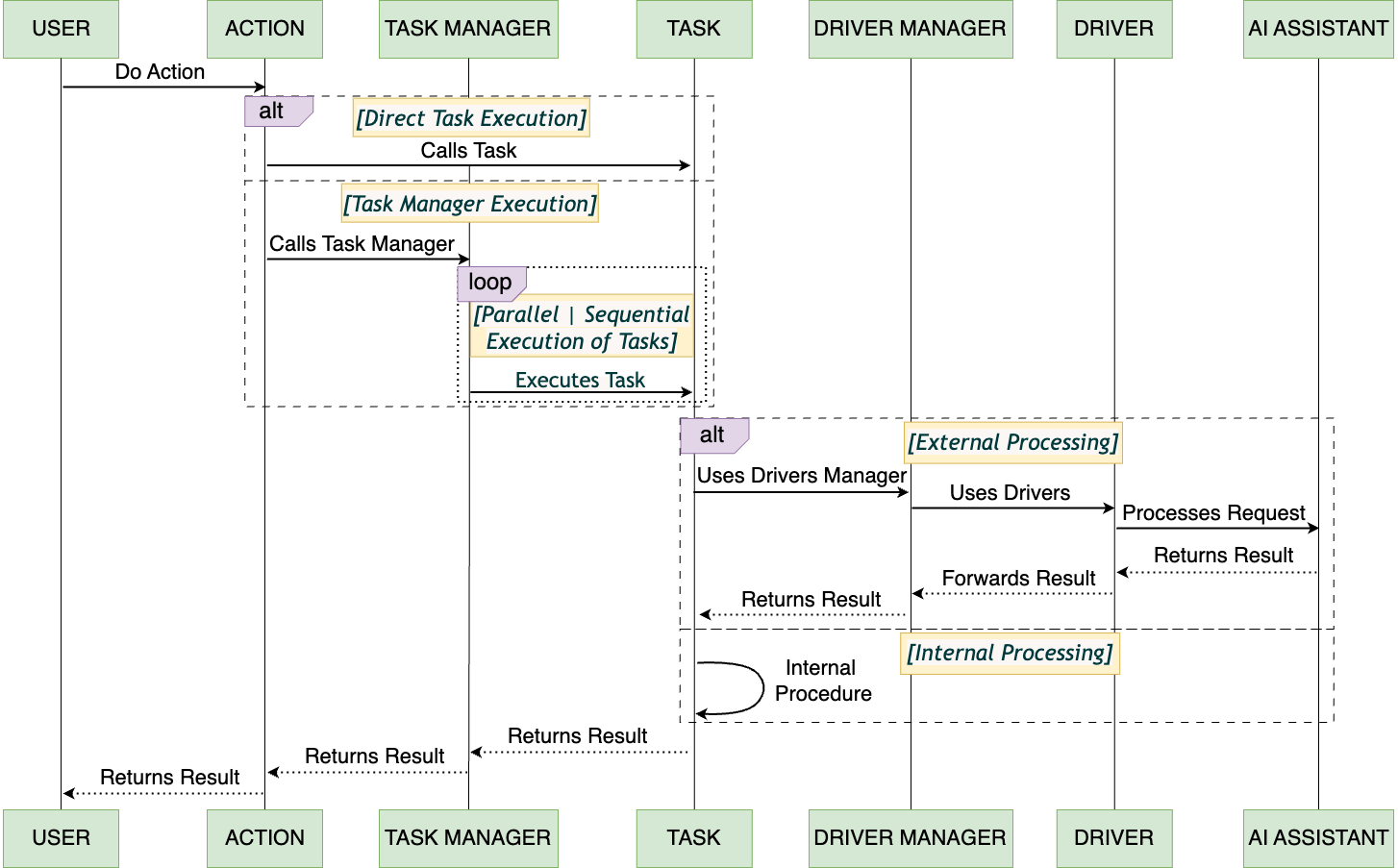}
        \caption{Sequence diagram showing the execution flow from the moment the user invokes an Action to the time the result is returned.}
        \label{fig:sequence} 
\end{figure*}

Figure~\ref{fig:sequence} illustrates the overall behavior of the \approach Plugin, from the generation of a user input to the visualization of the result.

The process starts with the \emph{user} executing a GUI event that initiates an \emph{Action}.

Depending on the complexity of the operation that must be executed the action may execute a \emph{task} directly or a \emph{task manager} to orchestrate the execution of multiple tasks.


A \emph{task manager} executes the tasks either in parallel or sequentially.

Each \emph{task} invokes external AI assistants and services as needed by the specific task. In some cases, a task may also implement a standalone procedure that does not require any interaction with AIs.

The interaction with AIs is mediated by the \emph{drives manager}, which acts as an intermediary for AI requests. For each individual request, the driver manager selects the appropriate \emph{driver}, which formats the request and forwards it to the selected \emph{AI Assistant}. 

The AI assistant processes the request and returns a response, which is propagated back through the driver to the driver manager and then to the task.

The task manager consolidates the results from multiple tasks and make the final result accessible to the action. 

The action delivers the final output to the User via the VSCode interface, displaying the AI-generated response, modifying the artefacts available in the IDE, or taking any other action.


\section{How to Implement AI-Assisted Analyses with \approach} \label{sec:Implementing AI-Assisted Analyses with Pincopallo}
In this section, we briefly discuss how a researcher or practitioners can exploit the \approach plugin to implement and experiment custom AI-mediated analyses by implementing three main elements: an action, a task manager, and a task. 

Once the user has identified the visual elements and the GUI events that must trigger the new AI-mediated operation, a corresponding \emph{action class} must be implemented. The action is configured to listen for the identified GUI events, and a callback method is invoked when these events are fired. The action class delegates the responsibility of executing the required task to either a task or a task manager. The action class is finally responsible for the visualization of the results once available. 

If a task manager is needed to execute the operation, it is possible to reuse the ready available \emph{task managers} that can orchestrate the parallel or sequential execution of one or more tasks. If a different execution protocol is needed, the available task managers can be extended and customized by the user.

The specific \emph{task} that must be accomplished by interacting with an AI assistant is implemented within a task. In many cases, the implementation consists of a carefully crafted prompt. 

For the interaction with AI assistants, the available \emph{drivers} can be reused. If necessary, a new driver can be added to the plugin to interact with new assistants. Each driver has an id that can be used by tasks to direct requests to specific driver instances.

In a nutshell, a researcher or a practitioner can quickly and easily implement and experiment with a new AI-mediated operation by (i) creating a new action, (ii) reusing or extending a task manager, and (iii) implementing a prompt in a task.

Our plugin is publicly available for reuse or extension at \repository.

\section{\approach In Action} \label{sec:IDENAMEInAction}
\begin{figure*}[ht]
        \centering
        \includegraphics[width=0.98\textwidth]{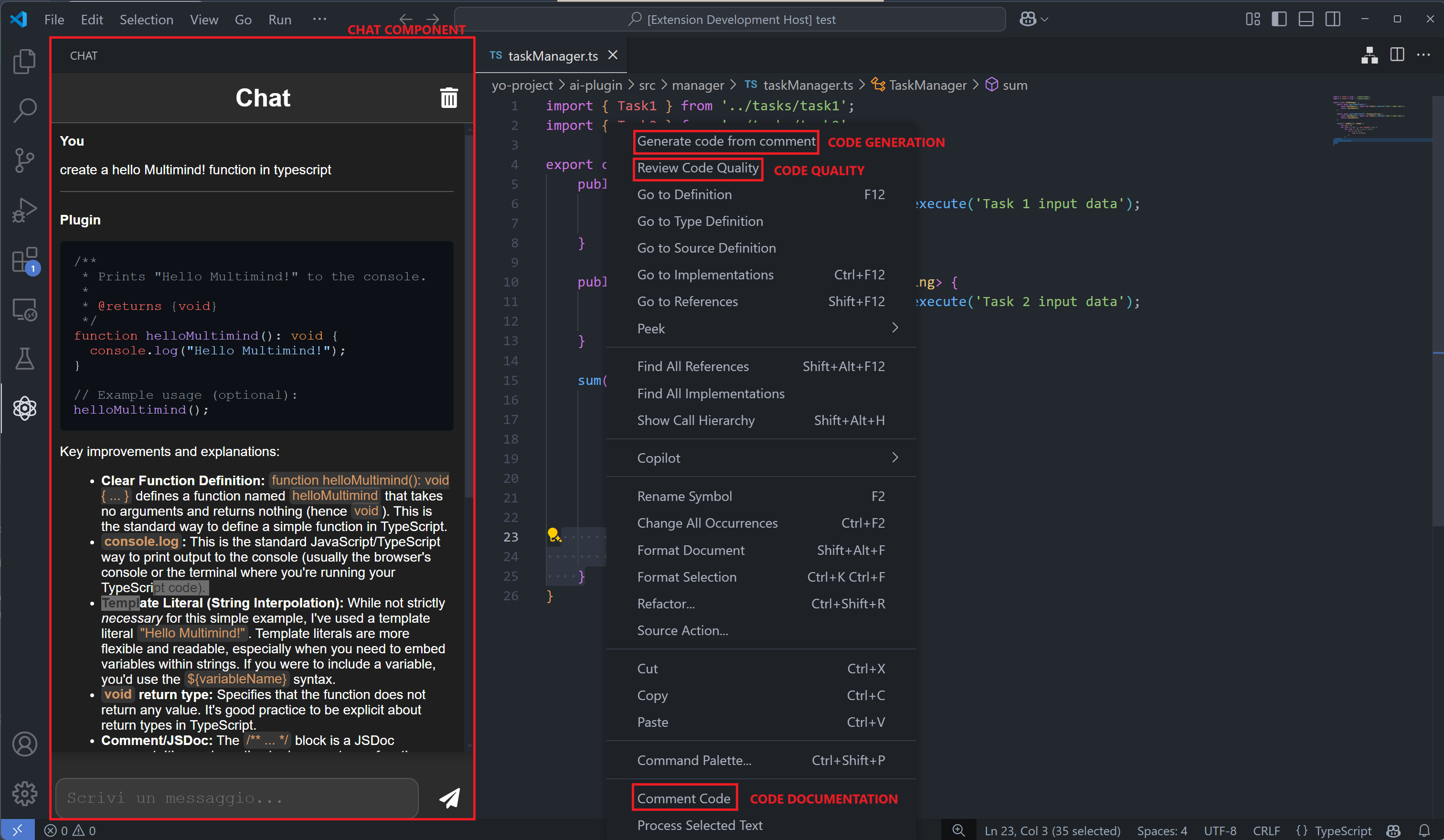}
        \caption{VSCode augmented with Multimind and the four exemplary analyses.}
        \label{fig:multimindScreen} 
\end{figure*}

To demonstrate the capabilities of \approach, we developed four exemplary analyses available in our repository: AI-Assisted Documentation Generation, which exploits the collaboration among multiple LLMs to generate code documentation, AI Interactive Chat, which allows textual interaction with AI assistants, Code Generation, which can be used to automatically generate the implementation of methods, and Documentation Review, which can be used to improve code documentation. These analyses are activated through additional menus and views, as shown in Figure~\ref{fig:multimindScreen}.
Due to space constraints, we will focus our discussion on the first two analyses. Additionally, we will highlight emerging research opportunities enabled by \approach that we are currently exploring.

\subsection{AI-Assisted Documentation Generation}

With this implementation, we address the use case of a developer who wants to exploit the capabilities of AI assistants to automatically generate source code documentation from code (e.g., Javadoc), enabling developers to cost-effectively obtain well-documented versions of their code through AI-assisted processing.

The comment generation task is started by selecting a portion of code within the VSCode editor. By right-clicking on the selected code and choosing the "\textit{Comment Code}" option from the context menu, the system initiates the documentation generation workflow. The triggered Action creates a request for the code generation Task Manager, embedding information about the programming language in the request. 

Using AI, the task manager runs two sequential tasks: the\linebreak \texttt{CommentTask}, which generates the code comments for the selected code, and the \texttt{DocQualityTask}, which determines if the code is of good quality or not. If the generated documentation is not satisfactory, the \texttt{CommentTask} is re-executed to regenerate the documentation. This iterative process continues until either the \texttt{DocQualityTask} confirms the adequacy of the generated comments or a predefined maximum number of iterations is achieved. If the maximum number of iterations is reached without achieving satisfactory documentation, the task manager stops the process and indicates that the documentation needs manual review or adjustment.

Two independent AI assistants are used to complete these tasks. In particular, the \textbf{OpenAI Driver} is used for generating code comments, while the \textbf{Gemini Driver} is used for assessing the quality of the generated comments.


The final output is rendered by the Action within the VSCode user interface, providing the developer with AI-enhanced documentation that is directly integrated and added to the code in their coding environment.

\subsection{Integration of a Chat Interface in the VSCode extension for AI Assistance}
A dedicated chat interface has been integrated into the VSCode extension to provide AI-driven assistance without disrupting the coding workflow. This feature allows developers to interact with AI assistants through a general-purpose communication channel, independent of inline code commenting.

The chat interface is accessible via an icon in the VSCode activity bar. When clicked, it opens a sidebar panel, ensuring direct access to an AI assistance. The interface is designed to be intuitive and responsive, facilitating seamless interaction.

The chat interface is managed by the \texttt{ChatViewProvider}, responsible for rendering and controlling the React-based frontend embedded within the VSCode sidebar. It utilizes \texttt{useState} and \texttt{useEffect} hooks to handle dynamic state updates and ensure real-time responsiveness.

Integration with the backend is established through a bidirectional channel 
that enables communication between the React frontend and the TypeScript backend. Messages from the developer are sent as VScode post messages, 
while AI responses are received through 
event listeners, ensuring a fluid and interactive experience.

The chat component processes \texttt{Open Task} requests, designed for context-free queries. These requests are forwarded to the\linebreak \texttt{DriverManager}, which selects and interacts with the appropriate AI assistants (e.g., OpenAI, Gemini). Multiple AI assistants process the query independently, and their responses are collected asynchronously.

Developers can compare the different responses and select the most suitable one, enhancing flexibility and efficiency in AI-assisted development. This multi-assistant approach provides diverse insights, improving decision-making and problem-solving within the development workflow.

\subsection{Research Playground}

\approach, in addition to offering a platform for the implementation of new analyses to be integrated with the IDE, is also a platform to experiment with new research approaches. For instance, we are currently exploiting the plugin to experiment two new research directions.

\emph{Learning Assistants}. We are currently working on introducing learning components within the IDEs that can monitor the interactions between developers and AI, and learn how to improve these interactions. In particular, learning could be used to discover \emph{how} and \emph{when} AI is triggered, and improve its seamless integration according to these patterns (e.g., not activating unwanted interactions, and timely activating desired AI recommendations). The learning component has to run within the IDE, mediating interactions between users and AI.

\emph{Personalized AI}. Effective AI interaction is not only about correctness, but also about providing personalized recommendations that match the requirements of the organization, the project, and the developers that use the AI. We are currently working on exploiting contextual information present within the IDE, and also present outside the IDE but accessible from it (e.g., information on GitHub), to generate contextualized recommendations. For instance, recommending code that matches the style used to implement the rest of the code present in the project, it has to be integrated in. Naturally, offering personalized AI assistants is not only about exploiting the context, but also about learning how to perform optimally for a given user.

\section{Related Work} \label{sec:RelatedWork}
The integration of LLMs within IDEs has gained significant attention in recent years, with a focus on leveraging these advanced tools to enhance developer productivity and understanding \cite{Nam:AICodeUnderstanding:ICSE:2024,Pinto:StackSpot:CAIN:2024,Ross:Programmer:IUI:2023,Chen:VSCuda:SC:2023,Sergeyuk:AIIDEReview:IDE:2024}.

Nam et al. \cite{Nam:AICodeUnderstanding:ICSE:2024}  investigate the role of an LLM-based conversational user interface within an IDE, demonstrating its potential to assist developers in understanding code by directly querying the language model with contextualized information. The findings prove that using such tools can significantly improve task completion rates compared to traditional methods, such as web searches for API documentation. This highlights the value of incorporating contextualization in LLM interactions to deliver relevant information seamlessly. This emphasis on contextualized interactions aligns with the goals of our work, which also seeks to optimize the developer experience by seamlessly integrating an LLM that responds to user queries based on the specific coding context supporting a wide range of developer needs.

Pinto et al. \cite{Pinto:StackSpot:CAIN:2024} explore developers’ information-seeking behaviors and identify challenges programmers face when utilizing AI assistants for coding tasks. Their results emphasize the importance of designing user-friendly tools that simplify interactions with AI. In line with this finding, our plugin can be used to minimize cognitive load through a prompt-less interaction mode, as early experimented with the comment generation analysis, allowing developers to obtain relevant information directly within the IDE. \approach aims to reduce the barriers developers encounter when seeking assistance in their software development activities, enabling more intuitive and efficient collaboration between developers and multiple AI assistants.

The Programmer’s Assistant \cite{Ross:Programmer:IUI:2023} emphasizes multi-turn conversational interactions with large language models, allowing developers to refine queries and engage in productive dialogues about coding tasks. This approach enhances debugging and learning through context retention. In contrast, \approach focuses on modularity and integration of multiple AI assistants, enabling developers to customize interactions for various tasks. While The Programmer's Assistant prioritizes conversational context, \approach allows for a wider range of AI capabilities, streamlining diverse workflows within the IDE.

Chen et al. \cite{Chen:VSCuda:SC:2023} introduce VSCuda, a targeted approach to enhancing CUDA programming productivity by integrating LLMs within the Visual Studio Code environment. VSCuda provides context-aware features such as code completion, syntax highlighting, and optimization suggestions specifically for CUDA, demonstrating how LLMs can deliver immediate assistance tailored to the programming domain. In contrast to our work, which aims for a modular integration of various AI assistants across multiple development tasks, VSCuda emphasizes specialized support for CUDA programming challenges. Both approaches highlight the importance of LLM integration in IDEs to improve the developer experience, with VSCuda focusing on domain-specific needs while \approach facilitates versatile interactions with multiple AI assistants.

While a growing number of VS Code extensions now incorporate AI code assistants or LLMs, most current solutions are primarily designed for end-users and offer limited extensibility or multi-agent coordination capabilities. Notable among these tools are \emph{GitHub Copilot} \cite{Copilot}, \emph{Qodo Gen} \cite{Qodo}, \emph{Codeium} \cite{Codeium}, \emph{Tabnine} \cite{Tabnine}, and \emph{Cody by Sourcegraph} \cite{CodyBySourcegraph}. These systems deliver intelligent code suggestions, NL interfaces, and automation capabilities; however, they largely operate as monolithic assistants, typically relying on a single LLM backend and offering little to no support for orchestrating tasks across multiple AI models or for conducting empirical research.

In contrast, MultiMind has been conceived as an open and modular framework to support the development, customization, and empirical evaluation of AI-assisted tasks within the IDE. MultiMind originally offers both open access to its source code and the explicit support for the integration and orchestration of multiple LLMs, as well as a composable architecture that separates concerns between user-triggered actions, tasks, and assistant-specific drivers. Furthermore, it is explicitly designed to facilitate the rapid prototyping and evaluation of novel AI-augmented development workflows, making it a valuable asset for both researchers and advanced practitioners seeking to experiment with next-generation IDE functionalities.

In summary, while existing research demonstrates the potential of LLM technologies to improve various aspects of software development, there remains an urgent need for further investigation into designing task-specific user interfaces, simplifying the integration of AI into development workflows, managing interactions with multiple AI assistants, and improving the usability of AI-generated outputs~\cite{Sergeyuk:AIIDEReview:IDE:2024}. \approach addresses these challenges by providing a modular and extensible framework that enables developers to implement and experiment with new AI-assisted development tasks seamlessly. By facilitating interactions with various AI assistants, \approach promotes efficient collaboration in software development activities.

\section{Conclusion} \label{sec:Conclusion}
Studying how AI assistants and human developers can collaborate in development tasks is an emerging and important research field. Indeed, the IDE is one of the most relevant places where this collaboration happens. In order to experiment novel approaches and interaction modes, it is necessary to have access to tools and environments that facilitate their implementation and experimentation.

In this paper, we introduced \approach, a Visual Studio Code plugin designed to facilitate AI-assisted development tasks by enabling seamless integration with multiple AI assistants. Our framework provides a modular and extensible environment, allowing developers and researchers to experiment with new AI-driven interactions without the need for extensive modifications to their IDE environment. 

The results of our implementation and the example cases implemented demonstrate the effectiveness of \approach in supporting AI-assisted documentation generation, code generation, and interactive chat-based development support. The architecture of \approach ensures flexibility, enabling both structured and open-ended interactions with AI assistants, which can be tailored to different development needs.

Despite these promising contributions, several challenges remain open for future research. Enhancing the adaptability of AI interactions based on user behavior, optimizing AI-driven feedback loops, and improving the personalization of AI recommendations are key areas that require further investigation. Additionally, the integration of more advanced learning mechanisms within the IDE to refine AI suggestions over time could further improve the efficiency and effectiveness of AI-assisted development workflows.

We believe that \approach represents an initial step towards AI-augmented software development, providing a foundation for further research and practical adoption of AI assistants within modern IDEs. Future work will focus on extending the capabilities of \approach by incorporating more sophisticated task orchestration mechanisms, expanding its support for additional AI assistants, and conducting extensive empirical evaluations to measure its impact on developer productivity.

\begin{acks}
This work was partially supported by the MUR under the grant ``Dipartimenti di Eccellenza 2023-2027'' of the Department of Informatics, Systems and Communication of the University of Milano-Bicocca, Italy; and by the Engineered MachinE Learning-intensive IoT systems (EMELIOT) national research project, which has been funded by the MUR under the PRIN 2020 program\linebreak (Contract 2020W3A5FY).
\end{acks}

\balance 

\bibliographystyle{ACM-Reference-Format}
\bibliography{main}

\end{document}